\documentclass[sn-nature]{sn-jnl}

\usepackage{graphicx}%
\usepackage{multirow}%
\usepackage{amsmath,amssymb,amsfonts}%
\usepackage{amsthm}%
\usepackage{mathrsfs}%
\usepackage[title]{appendix}%
\usepackage{xcolor}%
\usepackage{textcomp}%
\usepackage{manyfoot}%
\usepackage{booktabs}%
\usepackage{algorithm}%
\usepackage{algorithmicx}%
\usepackage{algpseudocode}%
\usepackage{listings}%

\theoremstyle{thmstyleone}%
\theoremstyle{thmstyletwo}%

\theoremstyle{thmstylethree}%

\begin{document}

\title[Article Title]{All-dielectric meta-waveguides for flexible polarization control of guided light }


\author[1]{\fnm{Syuzanna} \sur{Asadulina}}

\author[2]{\fnm{Andrey} \sur{Bogdanov}}\email{a.bogdanov@hrbeu.edu.cn}

\author[1]{\fnm{Stanislav} \sur{Glybovski}}

\author[1]{\fnm{Oleh} \sur{Yermakov}}\email{oe.yermakov@gmail.com}

\affil[1]{Independent Researcher}

\affil[2]{\orgdiv{Qingdao Innovation and Development Center}, \orgname{Harbin Engineering University}, \orgaddress{\city{Shandong}, \country{China}}}



\abstract{Guided waves are the perfect carriers of electromagnetic signals in planar miniaturized devices due to their high localization and controlled propagation direction. However, it is still a challenge  to control the polarization of propagating guided waves. In this work, we discover both theoretically and experimentally the broadband polarization TE-TM degeneracy of highly localized guided waves propagating along an all-dielectric metasurface and a subwavelength chain of dielectric high-index cylinders. Using the discovered near-field polarization degree of freedom, we demonstrate experimentally the polarization transformation for guided waves propagating along a subwavelength chain of dielectric cylinders at any frequency within the finite spectral range. Namely, we implement the simplest planar near-field polarization device -- the quarter-wave-retardation (linear-to-circular) polarization transformer of guided waves. The results obtained discover the polarization degree of freedom for guided waves paving the way towards numerous applications in flat optics and planar photonics.}




\maketitle

\newpage
\section{Introduction}\label{sec1}

Polarization represents an electromagnetic unit of information for transferring and encoding the optical signal forming a base of optical cryptography and security for communication systems~\cite{javidi2000polarization,tang2014experimental,ding2021recent}. It may be easily controlled and adjusted for plane waves in 3D medium via bulk anisotropic wave plates~\cite{born2013principles} or ultrathin planar metasurfaces~\cite{arbabi2015dielectric,glybovski2016metasurfaces,kruk2016invited,baena2017broadband,mueller2017metasurface}. The conventional way to manipulate the polarization state of propagating waves is based on lifting off the degeneracy between two polarizations and further control the phase delay between them.
The polarization degeneracy allows for the preservation of the polarization state during the wave propagation, while the specified phase delay between propagating modes with orthogonal polarization leads to on-demand polarization engineering. The polarization degeneracy is the inherent feature of the bulk isotropic medium. It means a complete coincidence in the dispersion of TE- and TM-polarized (s- and p-polarized) plane waves. In other words, both TE and TM modes have the same propagation constants at any frequency and along any propagation direction, so their superposition may lead to any polarization state.

Anisotropy of the medium or even presence of interfaces results in the breaking of the degeneracy of TE and TM modes, which is supported by different Fresnel coefficients for s- and p-polarized waves. The typical examples of structures, where TE and TM modes are non-degenerate, include uni- and bi-axial crystals, isotropic parallel-plate waveguides, surface waves, etc. As a consequence, planar photonic devices do not possess the polarization degree of freedom for light, which significantly limits the range of applications. 

In the near-field spectrum, i.e. for localized electromagnetic waves, the polarization TE-TM degeneracy is generally broken making it impossible to control the polarization of a localized wave during its propagation. One exception is the waveguide with symmetric cross-section (typically, circular or rectangular waveguides), which is independent of rotation by 90$^\circ$ around its axis. So, it is possible to control the guided-wave polarization inside such a waveguide using the interference of two modes of the same polarization (either TE or TM), but having orthogonal directions of transverse electric field vectors. It has been recently implemented as the standing guided waves inside the nanofiber with a polarization controlled by the couplers~\cite{joos2019complete,lei2019complete}. This implementation may potentially support the guided wave propagation along only one direction (along the waveguide or nanofiber) and requires a specific complicated coupling scheme. Another exception is the accidental polarization TE-TM degeneracy of surface and guided waves taking place due to the anisotropy-induced intersections of dispersion curves~\cite{yermakov2016spin}. This implementation allows to achieve TE-TM polarization degeneracy only at some single frequencies, while their group velocities are still different and even small frequency detuning sharply breaks the TE-TM degeneracy significantly limiting the practical applications. 

The main idea of the polarization degeneracy, implemented in this work, is based on the overlapping of the effective electric and magnetic resonant-like response of the structure in the near-field~\cite{kuznetsov2016optically}. It has been recently shown that dielectric particles exhibit electric and magnetic dipole-like and higher-multipolar Mie resonances~\cite{evlyukhin2012demonstration,kuznetsov2012magnetic}. The interference between electric and magnetic resonances in all-dielectric metasurfaces leads to a number of exciting phenomena highlighting the novel and promising field of meta-optics~\cite{kruk2017functional,camayd2020multifunctional,kivshar2022rise}. In particular, we refer to Huygens' metasurfaces allowing to achieve almost purely forward scattering of plane waves with over 95\% of transmittance and tailored phase based on the collective lattice Kerker conditions~\cite{babicheva2017resonant}, i.e. the overlapping of electric- and magnetic-like dipolar resonances in the reflection or transmission spectra~\cite{decker2015high,chen2018huygens}. However, the TE and TM guided modes of Huygens' metasurface are still non-degenerate which is a consequence of the spatial dispersion~\cite{bobylev2020nonlocal}. Nevertheless, being motivated by the independently excited electric and magnetic contributions in Huygens' metasurfaces, we optimize the designs of all-dielectric periodic subwavelength structures in a way to achieve the overlapping of electric and magnetic eigenmodes in the near-field. In other words, we aim to fulfill the Kerker effect not for the plane waves impinging on a metasurface, but for the guided waves propagating along a metasurface or a meta-waveguide.

\begin{figure*}[t] 
    \centering
    \includegraphics[width=0.8\linewidth]{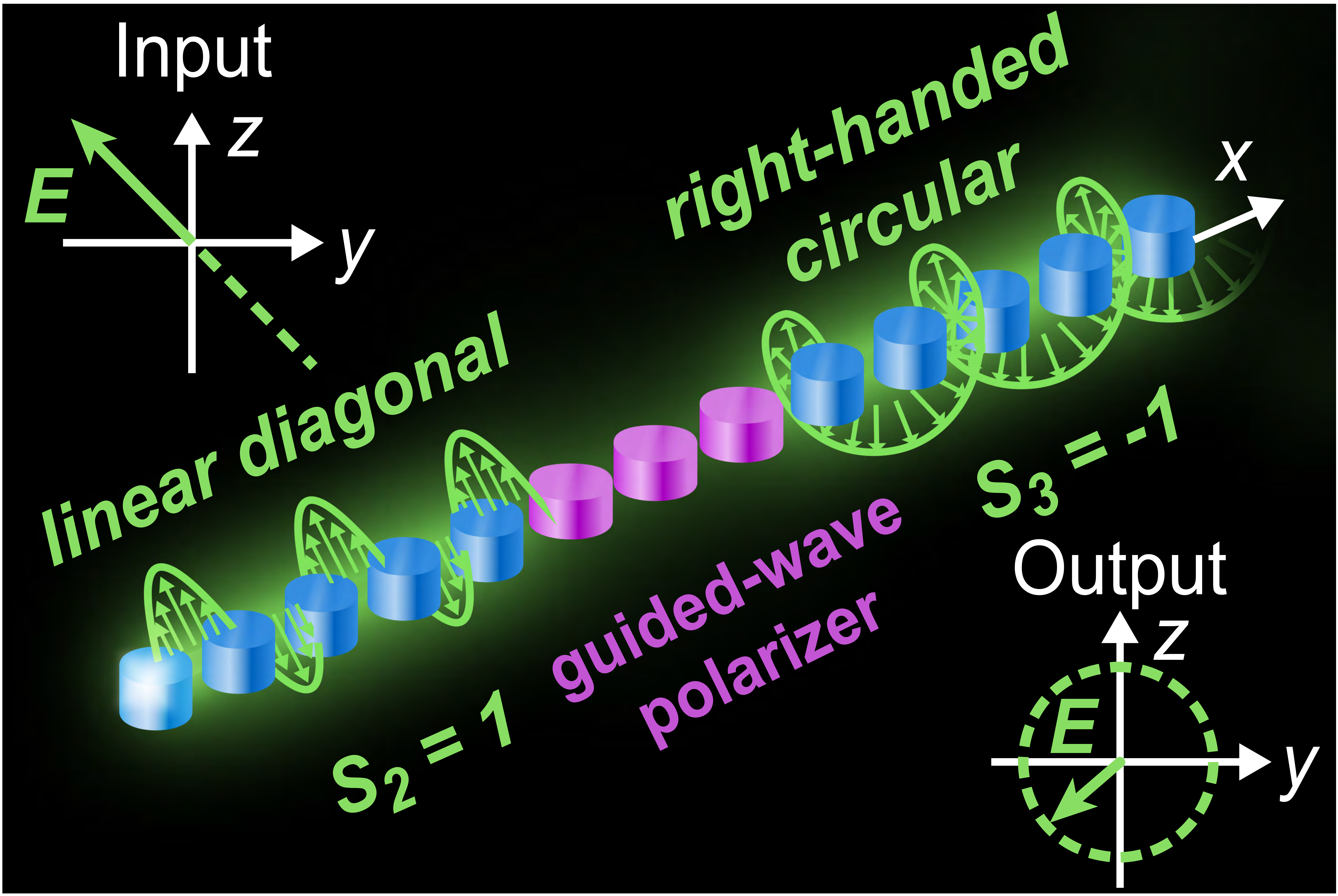}
    \caption{Conceptual sketch of the near-field planar linear-to-circular polarizer of guided waves based on the TE-TM degeneracy in all-dielectric 1D meta-waveguide representing a chain of cylinders. It transforms the excited linear diagonal polarization into the right-handed circular one.}
    \label{fig:concept}
\end{figure*}

In this work, we bring the polarization degree of freedom to the domain of guided waves that extend the polarization control of light to the planar two-dimensional photonic devices (Fig.~\ref{fig:concept}).
We propose an approach to engineering of all-dielectric nanostructures with degenerate TE-TM spectrum, which is quite general and it can be applied for the visible, infrared, THz, and microwave spectral ranges. Here, as a proof-of-concept experiment, we demonstrate the polarization TE-TM degeneracy of guided waves in the microwave. We investigate the guided modes of the 2D and 1D periodic subwavelength arrays of ceramic high-index cylinders representing 2D and 1D meta-waveguides, respectively. We provide the designs of two all-dielectric structures (a 2D and a 1D) with polarization-degenerate dispersions of TE- and TM-guided modes, and analyze the degree of the polarization degeneracy. The degenerate dispersion of TE and TM modes for a chain of ceramic cylinders has been verified in a microwave experiment (Fig.~\ref{fig:geom+setup}). Based on the discovered platform with TE-TM polarization-degenerate spectrum, we have numerically and experimentally demonstrated the functionality of a planar linear-to-circular polarization transformer for guided waves (Fig.~\ref{fig:concept}). The results obtained form the fundamentally novel platform for flat electromagnetic devices aimed to control polarization in the near-field.

\section{Results}\label{sec2}

\subsection{Definitions and optimization goals}

\subsubsection{Degree of polarization degeneracy} 

The keynote value of the dispersion engineering task under consideration is the degree of polarization degeneracy (DoPD). It evaluates how close the dispersions of TE and TM modes are and, as a consequence, characterizes the phase delay between them. First, we introduce the wavevector-related degree of polarization degeneracy ($k$-DoPD) as
\begin{equation}
    \Delta k = |k_{\text{TE}} - k_{\text{TM}}|,
\end{equation}
where $k_{\text{TE}}$ and $k_{\text{TM}}$ are the wavevector values of TE and TM modes at the same frequency, respectively. This is the most convenient value for the polarization degeneracy determination. For instance, it is well-known that $\Delta k L = \pi/2$, where $L$ is the length of the polarizer, means a quarter-wave retardation and leads to the linear-to-circular polarization conversion.

From another perspective, it is also relevant to introduce the absolute frequency-related degree of polarization degeneracy ($f$-DoPD) as
\begin{equation}
    \Delta f = |f_{\text{TE}} - f_{\text{TM}}|,
\end{equation}
where $f_{\text{TE}}$ and $f_{\text{TM}}$ are the frequencies of TE and TM modes at the same value of wavevector, respectively. This value is especially useful for the flat dispersion bands in the vicinity of resonances, where the $k$-DoPD may have large values despite the extremely low difference in the resonances of TE and TM modes. 

\begin{figure*}[t] 
    \centering
    \includegraphics[width=0.8\linewidth]{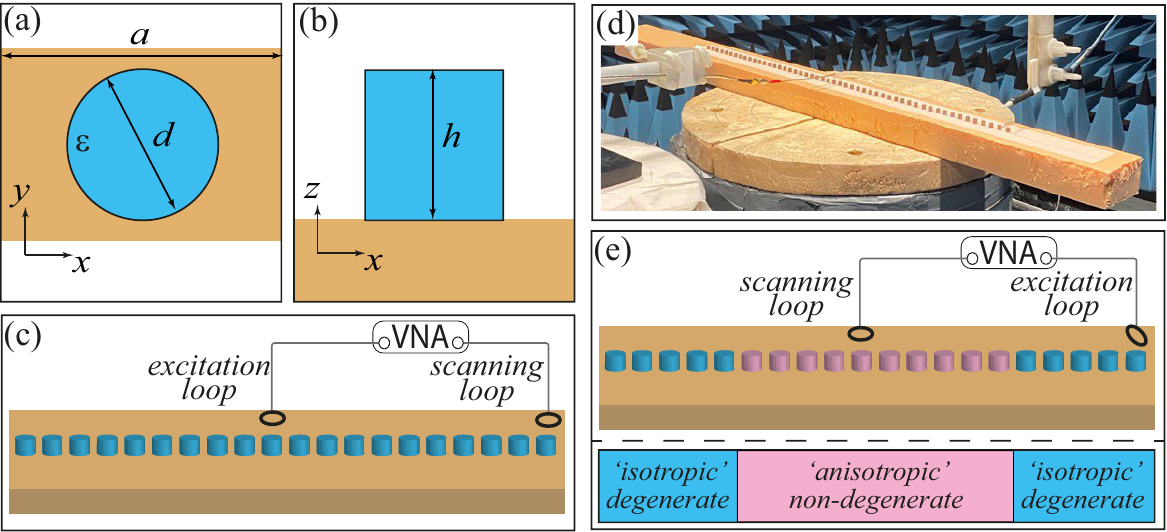}
    \caption{Sample and experimental setup. (a,b) Top (a) and side (b) view of a unit cell consisting of a ceramic ($\varepsilon$) cylinder with a diameter $d$ and height $h$ placed at the surface of foam holder. The period of the structure is $a$. (c,d) Sketch and photo of the experimental setup measuring the spatial distribution of normal component of the magnetic field in a chain of cylinders. The excitation loop is located above the central cylinder, the scanning loop is moving along a chain. (e) Experimental setup measuring the spatial distribution of normal component of the magnetic field in a planar polarizer. Planar polarizer represents a chain of 'isotropic' and 'anisotropic' cylinders supporting degenerate and non-degenerate eigenspectra, respectively. The excitation loop is located near the edge of the structure and rotated by $45^\circ$. Here, VNA is the vector network analyzer.}
    \label{fig:geom+setup}
\end{figure*}

\subsubsection{Degree of localization}

As we will show further, the DoPD is inversely proportional to the degree of localization (DoL) of the guided wave. Intuitively, it is clear because the dispersions of TE- and TM-polarized plane waves in a homogeneous isotropic medium are completely degenerate, but not localized. On the contrary, the TE- and TM-guided waves in a dielectric slab are strongly localized, but not degenerate. So, there is an interplay between DoPD and DoL. Therefore, we introduce the degree of localization as the ratio between the phase velocities of guided and plane waves:
\begin{equation}
    \Delta L = \frac{c}{\text{min}\,v_g} = \frac{k_B}{k_0} = \frac{c}{2 a f},
\end{equation}
where $\text{min}\,v_g = \omega/k_B$ is the minimum phase velocity of the guided wave, $k_B = \pi/a$ is wave vector at the boundary of the first Brillouin zone in the square periodic structure with lattice constant $a$, $k_0 = 2 \pi f/c$ is the wave vector of a plane wave in vacuum, $f$ is the operational frequency, $c$ is the speed of light. Thus, the DoL is defined as the ratio of the maximum possible wave vector of the guided wave to the corresponding wave vector of a plane wave at the same frequency.

\subsubsection{Stokes parameters \label{sec:Stokes-intro}}

In Section~\ref{sec:polarizer}, we analyze the polarization states of the propagating guided wave in the different cross-sections. For a better representation, we follow the Stokes parameters formalism~\cite{born2013principles}. Assuming the wave propagation along $x$-axis, the Stokes parameters are defined via $E_y$ and $E_z$ (alternatively, $H_y$ and $H_z$) field components  as follows
\begin{equation}
\begin{split}
    & S_0 = |E_{y}|^2 + |E_{z}|^2,\\
    & S_1 = |E_{y}|^2 - |E_{z}|^2, \\
    & S_2 = 2 |E_{y}| |E_{z}| \, \text{cos}\delta,\\
    & S_3 = 2 |E_{y}| |E_{z}| \, \text{sin}\delta,
\end{split}
\end{equation}
where $|E_{y,z}|$ are the amplitudes of $E_y$ and $E_z$ field components, and $\delta = \text{arg}E_y - \text{arg}E_z$ is the phase difference between these field components. Here, $S_0$ encodes the total intensity of electromagnetic field, $S_1$, $S_2$ and $S_3$ correspond to the degrees of horizontal (TE, $S_1 = 1$) and vertical (TM, $S_1 = -1$) linear, linear rotated by $45^\circ$ ($S_2 = 1$) and $-45^\circ$ ($S_2 = -1$) with respect to $y$-axis, and left-handed ($S_3 = 1$) and right-handed ($S_3 = -1$) circular polarizations, respectively. One can notice that $S_0^2=S_1^2+S_2^2+S_3^2$. Hereinafter, we normalized all Stokes parameters to $S_0$. 

The specific polarization state can be also marked at the Poincar{\'e} sphere drawn directly in three-dimensional Cartesian coordinates corresponding to ($S_1, S_2, S_3$) Stokes parameters. In the case of a fully polarized incident wave, the polarization state represents a point at the surface of the sphere of a radius $S_0$.

The linear-to-circular polarization transformation may be also analyzed via the total phase shift ($\Delta \delta$) acquired due to the polarizer. It may be defined as the difference ($\Delta \delta = \delta_a - \delta_b$) between the phase delays after ($\delta_a$) and before ($\delta_b$) polarizer.

Besides, we introduce the average Stokes parameters and total phase shift within a specified area  defined as follows:
\begin{equation}
\begin{split}
    & \overline{S}_{i} = \frac{\int{S_i \, dy dz}}{\int{dy dz}}, \; i = 1,2,3; \\
    & \overline{\Delta \delta} = \frac{\int{\Delta \delta \, dy dz}}{\int{dy dz}}.
    \end{split}
\end{equation}
The integrals are taken over the area $a\times a$ containing the cylinder cross-section.

\subsection{Polarization TE-TM degeneracy of guided modes}

\begin{figure*}[t]
\centering
\includegraphics[width=0.85\linewidth]{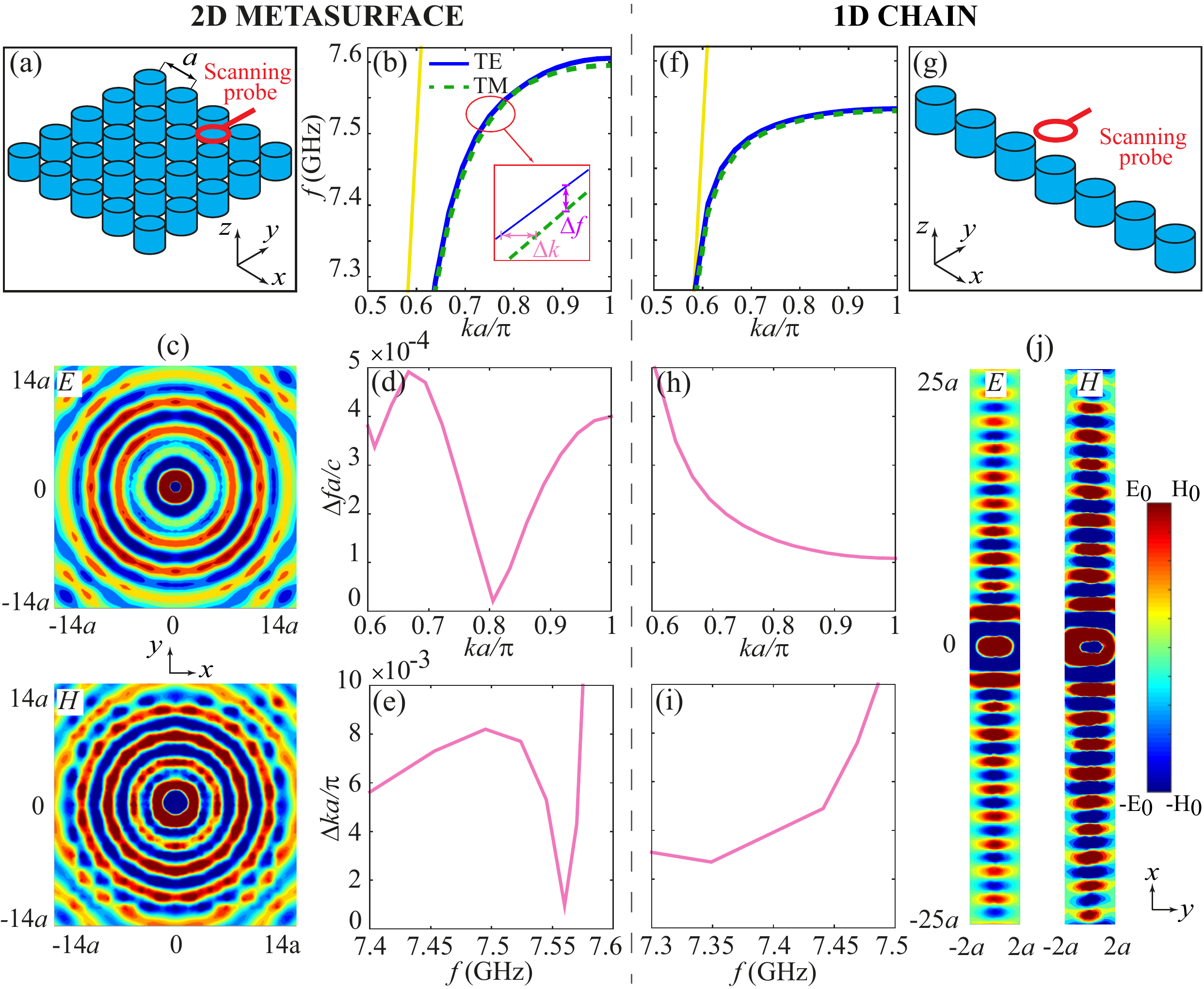}
\caption{(a,g) Sketch view of the considered periodic structures: (a) all-dielectric metasurface and (g) chain of 51 cylinders. (b,f) Numerically calculated dispersion curves of TE and TM localized waves, (d,h) $f$-DoPD and (e,i) $k$-DoPD for all-dielectric (a-e) 2D metasurface and (f-j) a chain of cylinders with the lattice constant $a$ surrounded by air. Yellow line in (b,f) corresponds to the light line in a vacuum. (c,j) Numerically calculated spatial distributions of the normal components of electric and magnetic fields for guided waves at the frequency $f= 7.5$~GHz in (c) 25$\times$25-cylinders all-dielectric metasurface and (j) a chain of 51 cylinders. The electric and magnetic fields were excited by a coaxial probe and loop, respectively, located at a distance $a/2$ above the central cylinder.}
\label{fig:2Dvs1D}
\end{figure*}

\begin{figure*}[t]
\centering
\includegraphics[width=0.8\linewidth]{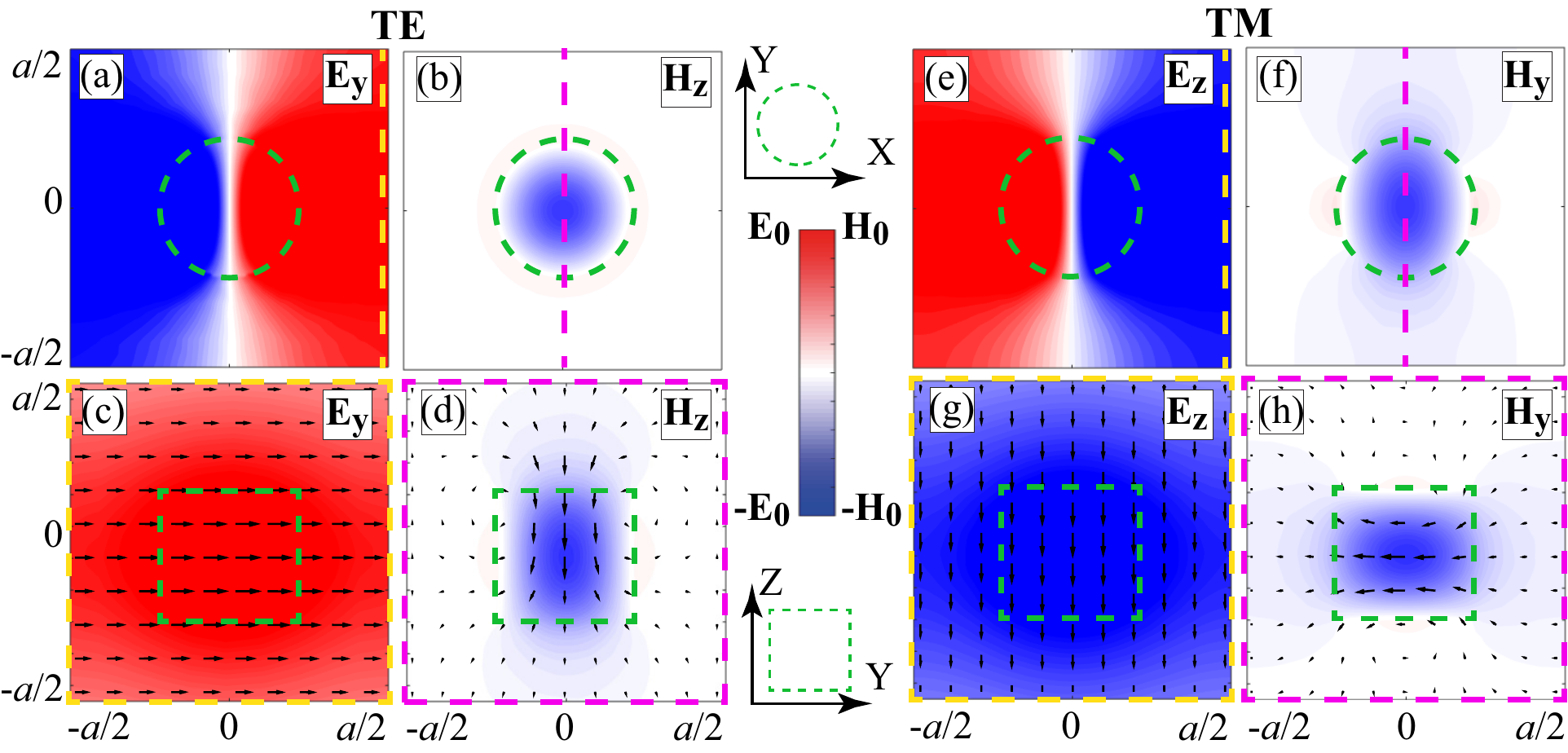}
\caption{Spatial distributions of TE (a-d) and TM (e-h) eigenmodes in $XY$- (a,b,e,f) and $YZ$-plane (c,d,g,h) at $f = 7.6$~GHz under $k = 0.95 \pi/a$. Namely, $E_y$ (a,c), $H_z$ (b,d) and $E_z$ (e,g), $H_y$ (f,h) components of TE and TM modes, respectively. The cross-section in $YZ$-plane is marked by the corresponding yellow and magenta dashed lines for electric (a,e) and magnetic (b,f) fields, respectively. The black arrows in (c,d,g,h) show the electric (c,g) and magnetic (d,h) field vectors at each point. The color bar shows the values of electric and magnetic fields, where $H_0 = 0.1 E_0$. The thin green dashed lines indicate the position of a ceramic cylinder.}
\label{fig:fields-eigen}
\end{figure*}

\begin{figure*}[t]
\centering
\includegraphics[width=0.75\linewidth]{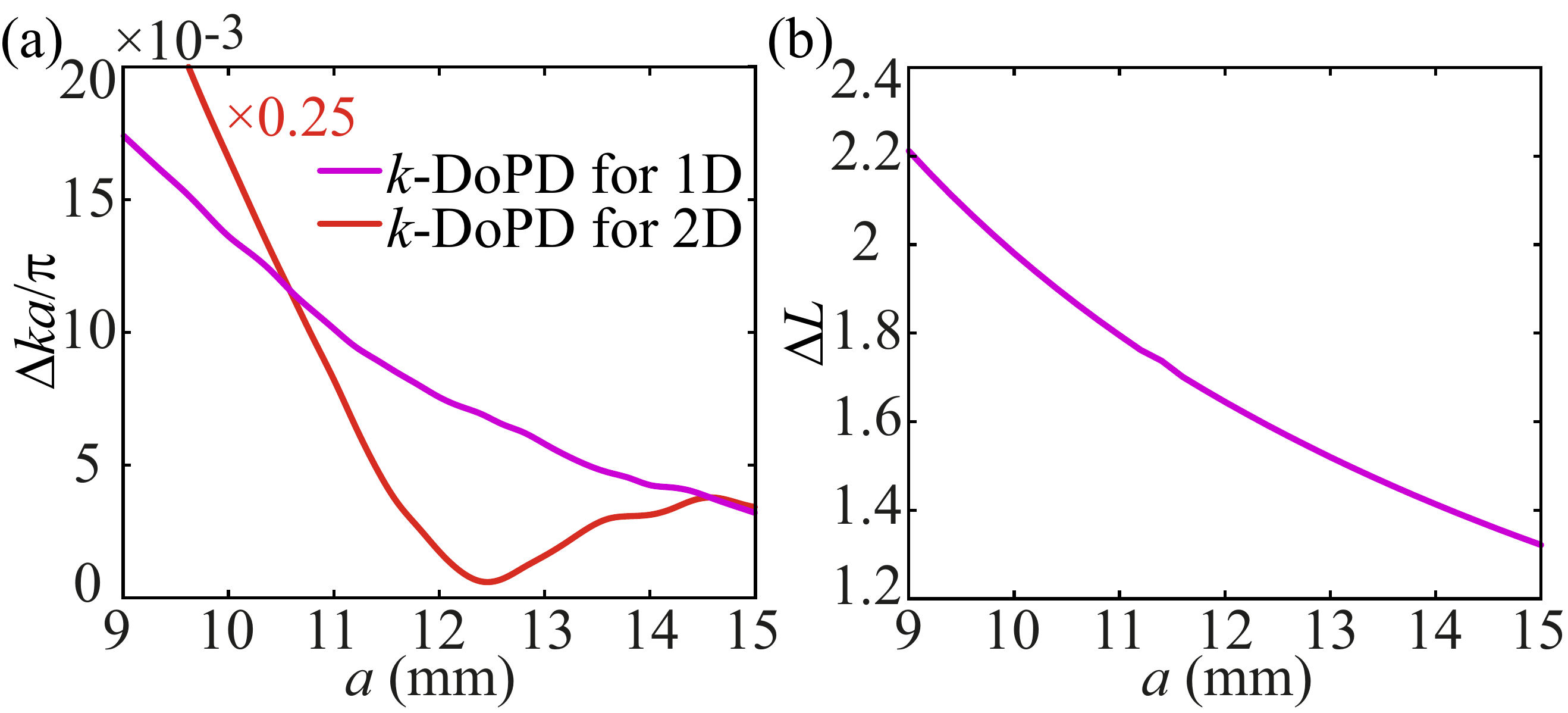}
\caption{Dependencies of (a) the maximum value of $k$-DoDP $\Delta k$ and (b) the DoL $\Delta L$ on the period $a$ for the 2D and 1D meta-waveguides based on the cylinders mentioned in Sections~\ref{sec:2D} and \ref{sec:chain}, respectively. The maximum values of $k$-DoDP for 2D meta-waveguide are reduced by four times. The DoL for 2D and 1D meta-waveguides differ within 1\%. }
\label{fig:degeneracy-vs-localization}
\end{figure*}

In this Section, we demonstrate both theoretically and experimentally the polarization-degenerate spectrum of guided waves. Namely, we consider the all-dielectric metasurface organized as a square 2D lattice of cylinders (Fig.~\ref{fig:2Dvs1D}a) and the 1D periodic chain of dielectric cylinders (Fig.~\ref{fig:2Dvs1D}g) as the two-dimensional and one-dimensional meta-waveguides, respectively. By fixing the diameter of the cylinder as $d = 5.2$~mm and its permittivity as $\varepsilon=45$, we optimize the height of the cylinders $h$ and a period of the structure $a$ in order to minimize the degree of polarization degeneracy in the range up to approximately 7.5~GHz (see Section~\ref{sec:num}). In both cases, we engineer the design and show that dispersions of TE- and TM-polarized guided modes are almost identical. Finally, we verify experimentally the TE-TM polarization-degenerate spectrum for a chain of cylinders by retrieving the dispersions of TE and TM modes from the near-field measurements (see Section~\ref{sec:exp}). This result discovers experimentally the near-field TE-TM polarization degree of freedom for the first time.   

\subsubsection{All-dielectric metasurface \label{sec:2D}}

The final design of the all-dielectric metasurface numerically optimized according to Section~\ref{sec:num} with $a=12$~mm and $h = 4.95$~mm leads to the well-degenerate eigenmodes spectrum in the range up to 7.57~GHz as it is shown in Fig.~\ref{fig:2Dvs1D}b. Figure~\ref{fig:fields-eigen} shows the field profiles of two eigenmodes propagating along $x$-axis in the vicinity of the cylinder, which allows to identify them as the mutually orthogonal TE and TM polarizations. The corresponding $f$-DoDP and $k$-DoDP are shown in Figs.~\ref{fig:2Dvs1D}d and \ref{fig:2Dvs1D}e, respectively. Namely, $\Delta f a /c < 5 \cdot 10^{-4}$ and $\Delta k a /\pi < 8 \cdot 10^{-3}$ within the frequency range up to 7.57~GHz. The sharp increase of $\Delta k$ at higher frequencies is associated with almost horizontal dispersion curves in the vicinity of the first Brillouin zone boundary (Fig.~\ref{fig:2Dvs1D}b). Finally, Fig.~\ref{fig:2Dvs1D}c shows the distributions of the normal components of electric ($E_z$) and magnetic ($H_z$) fields of guided waves at $25\times25$-cylinders structure excited by vertical electric dipole (electric source) and vertical magnetic dipole realized as a wire loop (magnetic source), respectively, at the frequency of 7.5~GHz.

\begin{figure*}[t]
\centering
\includegraphics[width=0.85\linewidth]{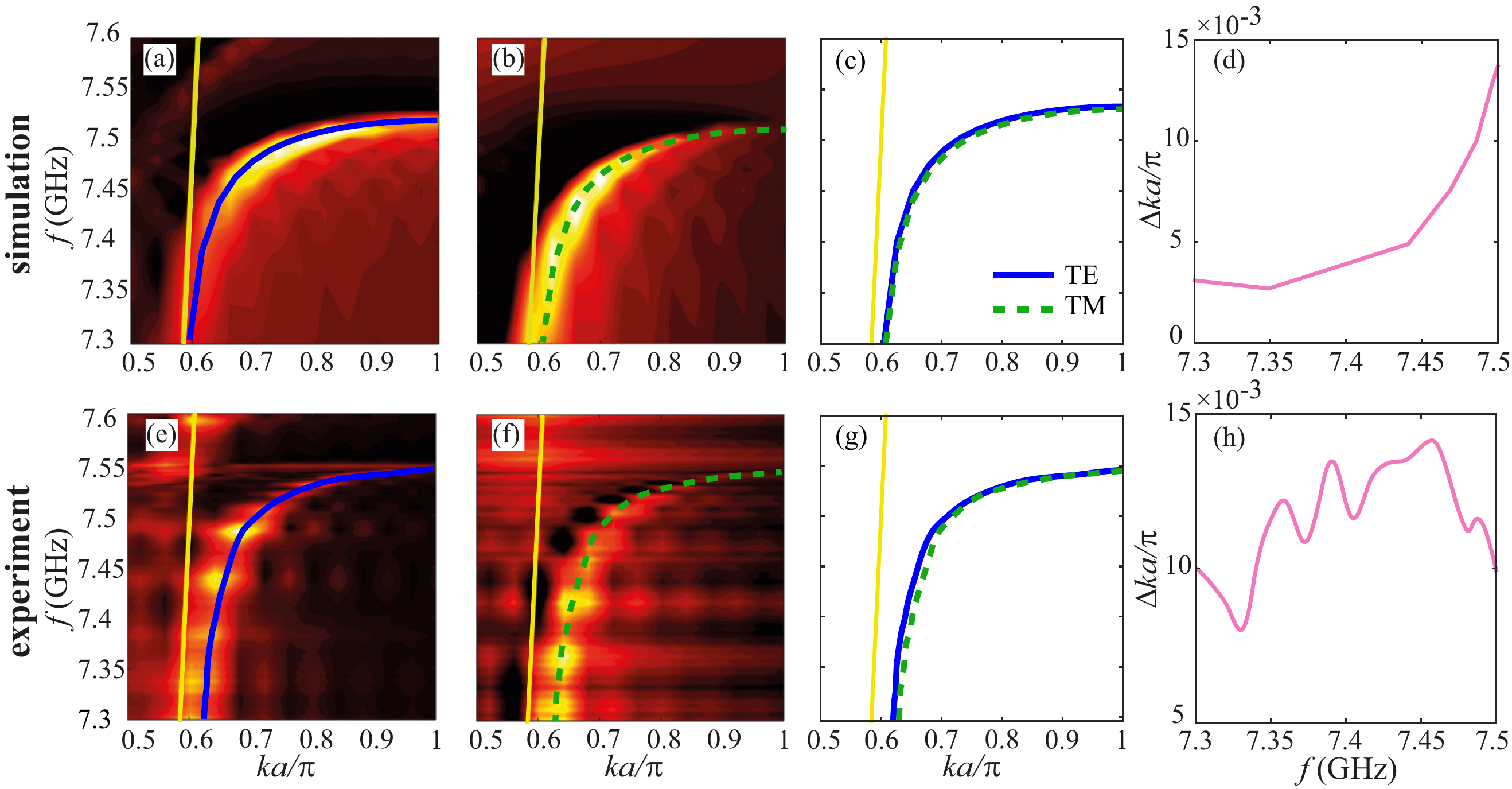}
\caption{(a,b,e,f) Dispersion curves of (a,e) TE and (b,f) TM modes restored from the normal components of electric and magnetic field distributions (a,b) calculated numerically within Eigenmodes solver of CST Microwave Studio and (e,f) measured experimentally. Solid blue and dashed green lines show the fitted local maxima of color maps and correspond to the dispersion lines of TE and TM modes, respectively. Yellow line corresponds to the light line in a vacuum. (c,g) The restored dispersion curves lines from (a,b,e,f) without color maps. (d,h) The wavevector-related degree of polarization degeneracy is restored from the  (d) numerical simulation and (h) experimental measurements.}
\label{fig:Figure3}
\end{figure*}  

\subsubsection{Chain of high-index cylinders \label{sec:chain}}

The $\Delta k$ and $\Delta L$ are inversely proportional to the period of the structure (Fig.~\ref{fig:degeneracy-vs-localization}). It means that the minimization of $k$-DoPD leads to the delocalization of the guided wave. Therefore, we choose a period of $a=12$~mm corresponding to $\Delta L = 1.66$ and $\Delta k a / \pi < 6 \cdot 10^{-3}$. The decay path of guided waves, defined as the distance over which the intensity of the evanescent field drops to $1/e$, reaches 2.4~mm in the vicinity of resonance in this case.

The final optimized design of the chain of cylinders with $h = 5.3$~mm leads to the well-degenerate eigenmodes spectrum in the range up to 7.45~GHz as it is shown in Fig.~\ref{fig:2Dvs1D}f. The field profiles of two eigenmodes are similar to the ones shown in the inset of Fig.~\ref{fig:fields-eigen}. The corresponding $f$-DoDP and $k$-DoDP are shown in Figs.~\ref{fig:2Dvs1D}h and \ref{fig:2Dvs1D}i, respectively. Namely, $\Delta f a /c < 2 \cdot 10^{-4}$ for $k > 0.75 \pi/a$ and $\Delta k a /\pi < 6 \cdot 10^{-3}$ within the frequency range up to 7.45~GHz. Figure~\ref{fig:2Dvs1D}j shows the distributions of the normal components of electric ($E_z$) and magnetic ($H_z$) fields of guided waves at 51-cylinders structure excited by vertical electric dipole (electric source) and vertical magnetic dipole realized as a wire loop (magnetic source), respectively, at the frequency of 7.5~GHz. 

We numerically calculated and experimentally measured the spatial distributions of the normal components of electric and magnetic fields for a finite structure consisting of 51 cylinders in the frequency range from 7.3~GHz to 7.6~GHz (see Section~\ref{sec:exp}). Then, by using the two-dimensional Fourier transform~\cite{dockrey2016direct,yang2017hyperbolic,yermakov2018experimental,yermakov2021surface} we restored the color maps corresponding to the dispersions of TE and TM eigenmodes (Figs.~\ref{fig:Figure3}a, \ref{fig:Figure3}b, \ref{fig:Figure3}e and \ref{fig:Figure3}f). Fitting the local maxima values at each frequency we derived the proper dispersion dependencies (Figs.~\ref{fig:Figure3}c and \ref{fig:Figure3}g). Finally, we obtained the degree of polarization degeneracy that does not exceed $\Delta k a / \pi < 0.015$ for both numerically simulated and experimentally measured data.

\subsection{\label{sec:level4} Planar polarization transformer of guided waves \label{sec:polarizer}}

\begin{figure*}
   \centering
    \includegraphics[width=0.85\textwidth]{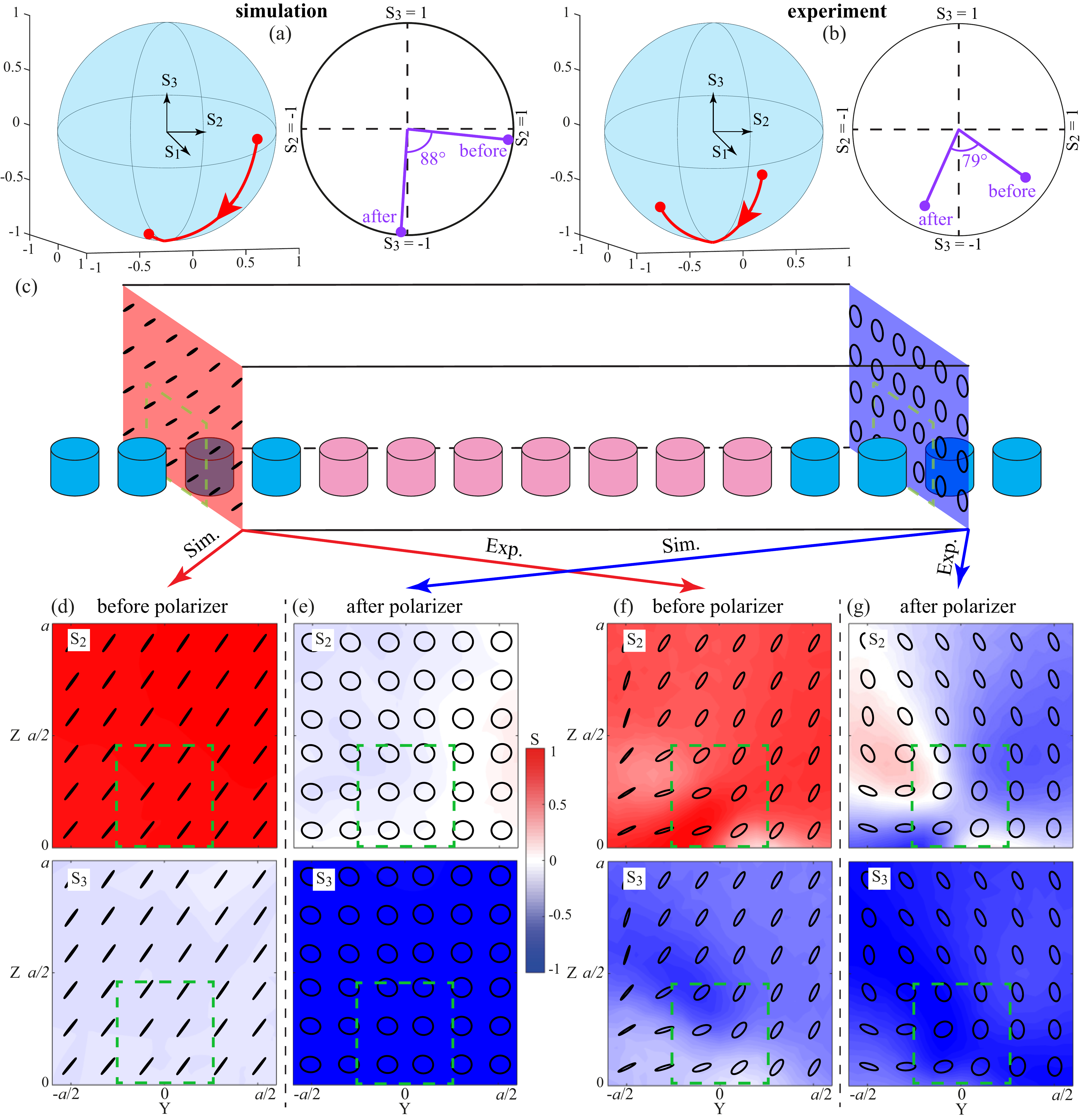}
    \caption{Stokes parameters before and after the planar polarizer. (a,b) Polarization states before and after the waveguide polarizer at Poincar{\'e} sphere of (a) calculated numerically in Frequency Domain Solver of CST Microwave Studio and (b) measured experimentally. Left and right panels show the trajectory of the polarization transformation and the corresponding cross-sections in $S_2-S_3$ plane, respectively. (d-g) The spatial distributions of $S_2$ and $S_3$ in $yz$-plane in the vicinity of cylinders (marked by the dashed green line) within the region $a \times a$ (d,e) calculated numerically and (f,g) measured experimentally (d,f) before and (e,g) after the polarizer. The black lines show the polarization states in special points. The spatial distributions of the fields were measured at a distance of two periods before and after the 'anisotropic' part depicted by pink cylinders in (c). }
    \label{fig:Stokes}
\end{figure*}

\begin{figure*}[t]
   \centering
    \includegraphics[width=0.85\linewidth]{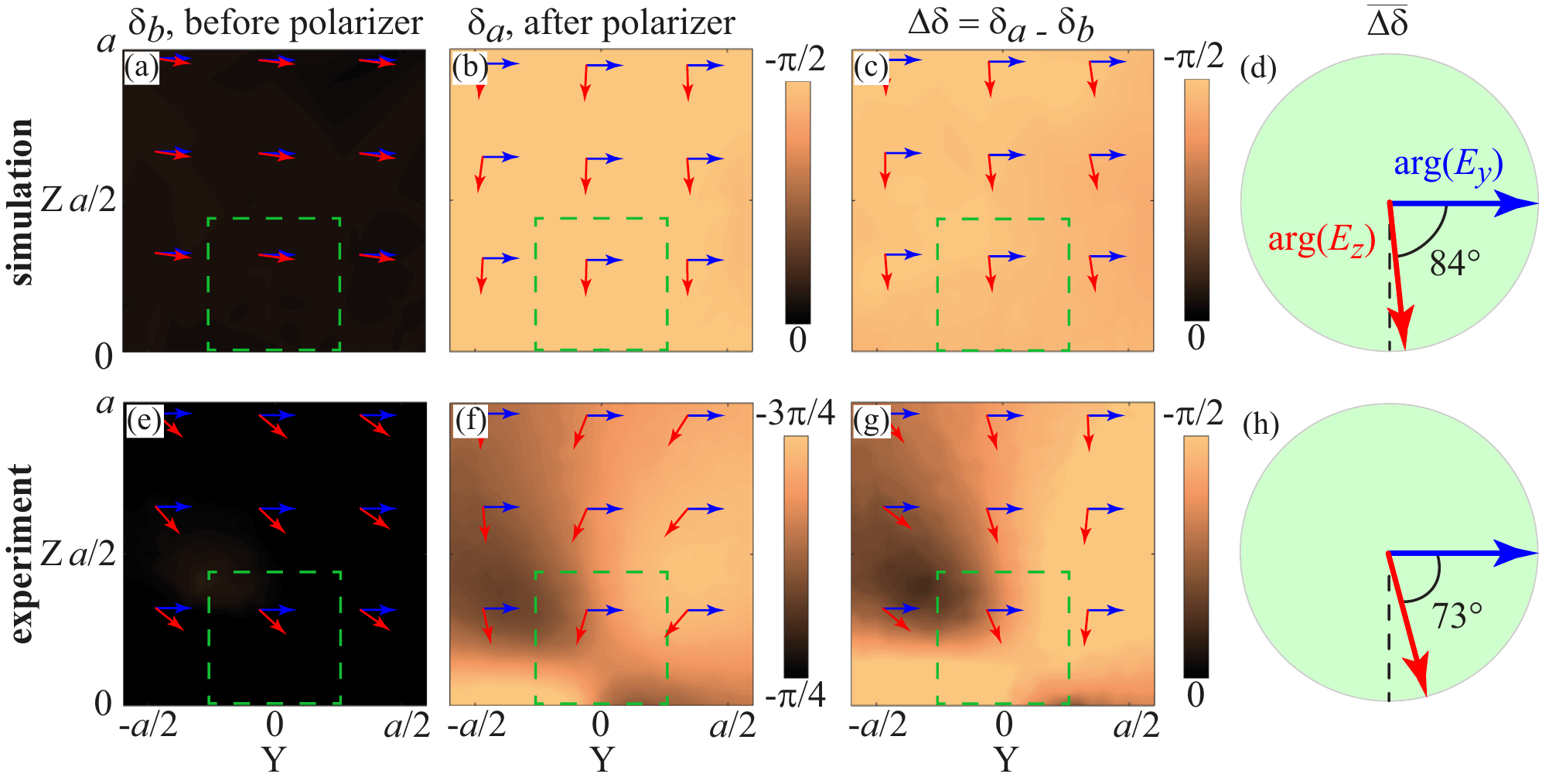}
    \caption{The spatial distributions of phase differences between $y-$ and $z-$ field components in $yz$-plane (a-d) calculated numerically and (e-h) measured experimentally (a,e) before and (b,f) after the polarizer. (c,g) The total phase shift calculated as the difference between the phase differences after and before the polarizer. Green dashed lines correspond to the arrangement of the cylinders. (d,h) The total phase shift averaged over the considered area. Blue and red arrows schematically show the phase of $y-$ and $z-$ field components. Color bars change from $0$ to $-\pi/2$ in (a-c,g) and from $-\pi/4$ to $-3\pi/4$ in (e,f).  }
    \label{fig:polariz-phase}
\end{figure*}

The structure of the polarizer consists of 3 parts 'isotropic'/'anisotropic'/'isotropic'. 'Isotropic' parts support the TE-TM degenerate guided waves ($\Delta k \approx 0$), while the 'anisotropic' part breaks the polarization degeneracy and brings the phase delay between two modes. The operational principle is completely analogous to the conventional wave plate, but it is applied for localized (guided) waves in the near-field instead of plane waves in the far-field. For the conventional wave plates, the roles of 'isotropic' and 'anisotropic' parts are played by the vacuum or isotropic dielectric medium and uniaxial crystal, respectively. 

The 'isotropic' part represents the chain of cylinders studied in Section~\ref{sec:chain}. The cylinders in the 'anisotropic' part have the diameter $d = 5.2$~mm and height $h = 5.4$~mm. The period is equal to $a=12$~mm within a chain. According to the numerical results, the 'isotropic' and 'anisotropic' parts are characterized by $\Delta k = 6 \cdot 10^{-3} \, \pi/a$ and $\Delta k = 5 \cdot 10^{-2} \, \pi/a$ at the frequency $f=7.45$~GHz, respectively.

The investigated planar polarizer is expected to bring a phase shift of $\pi/2$ between field components of guided waves at a propagation length of 108 mm. We consider it as a near-field analogue of a conventional quarter-wave plate. If the input polarization is linear and rotated at 45$^\circ$ with respect to $y$-axis, such a polarizer operates as a linear-to-circular (from $S_2 = \pm 1$ to $S_3 = \pm 1$) polarization transformer. 

The spatial distribution of the magnetic field in the $yz$-plane has been obtained numerically and experimentally following the methods described in Sections~\ref{sec:num} and \ref{sec:exp}. Then, we determined the spatial distributions of the Stokes parameters $S_2$ and $S_3$, and the total phase difference $\Delta \delta$ following the formalism mentioned in Section~\ref{sec:Stokes-intro}. To analyze the polarization transformation, we find the spatial distribution of the fields between cylinders in the $yz$-plane before and after the 'anisotropic' part, namely, retreating by two periods (Fig.~\ref{fig:Stokes}c). First, we present the spatial distributions of numerically calculated Stokes parameters $S_2$ and $S_3$ before (Fig.~\ref{fig:Stokes}d) and after (Fig.~\ref{fig:Stokes}e) the polarizer. In addition, we marked explicitly the polarization in 36 points with a step of 2~mm within the region $a\times a$ (12$\times$12~mm$^2$). One can notice the polarization transformation from the linear diagonal ($S_2 = 1$) to the right-handed circular ($S_3 = -1$) polarization (Figs.~\ref{fig:Stokes}d-\ref{fig:Stokes}e). Besides, we calculated the average Stokes parameters in the near-field and demonstrated the transformation from almost linear ($\overline{S}_2 = 0.94, \overline{S}_3 = -0.11$) to almost circular ($\overline{S}_2 = -0.06, \overline{S}_3 = -0.97$) polarization at the Poincar{\'e} sphere (Fig.~\ref{fig:Stokes}a). 

From other hand, we analyze the spatial distribution of phase difference $\delta$ between the $y$- and $z$-components of the field. We have shown that numerically simulated phase differences before (Fig.~\ref{fig:polariz-phase}a) and after (Fig.~\ref{fig:polariz-phase}b) polarizer are close to $0^\circ$ and $-90^\circ$ resulting in the total phase shift of about $-\pi/2$ (Figs.~\ref{fig:polariz-phase}c-\ref{fig:polariz-phase}d).

The numerical results were verified by the microwave experiment, see Section~\ref{sec:exp}. As a result, we restored the spatial distributions of $H_y$ and $H_z$ and found the Stokes parameters. One can notice the polarization transformation similar to the one from numerical simulation (Figs.~\ref{fig:Stokes}b, \ref{fig:Stokes}f and \ref{fig:Stokes}g). The polarization states of the guided wave before and after the polarizer are rather the ellipses elongated along the direction rotated by $45^\circ$ and $135^\circ$ with respect to the $y$-axis, respectively. The average Stokes parameters in the near-field change from $\overline{S}_2 = 0.62, \overline{S}_3 = -0.44$) before to $\overline{S}_2 = -0.32, \overline{S}_3 = -0.71$) after the polarizer, which is marked at the Poincar{\'e} sphere (Fig.~\ref{fig:Stokes}b). Although we did not excite the purely linear diagonal polarization at the input of the polarizer (due to the asymmetry of the coupling between the dipole source and the near-field of the propagating wave in the experiment), we have still demonstrated the polarization transformation with phase retardation of $\pi/2$ (Figs.~\ref{fig:polariz-phase}g-\ref{fig:polariz-phase}h). In this case, however, the measured phase differences before (Fig.~\ref{fig:polariz-phase}e) and after (Fig.~\ref{fig:polariz-phase}f) polarizer are close to $-\pi/4$  and $-3\pi/4$, which is in a good accordance with Figs.~\ref{fig:Stokes}f-\ref{fig:Stokes}g.

\section{Discussion}


We have demonstrated the degeneracy of TE and TM dispersion curves for 2D and 1D meta-waveguides composed of high-index ceramic cylinders. For the first time, we have experimentally shown the polarization-degenerate spectrum of TE and TM-guided waves in a subwavelength chain of cylinders, and evaluated the degree of polarization degeneracy.

The revealed broadband polarization TE-TM degeneracy of guided waves in all-dielectric waveguides opens a new degree of freedom for controlling localized light and extends the fundamental principles of polarization optics to the near-field. It was the missing brick for a number of applications requiring the polarization control of propagating localized light at any frequency within the finite spectral range. This result bridges the \textit{'polarization degree of freedom'} and \textit{'high localization'} inherent to the plane waves in a bulk isotropic medium and the guided waves in the miniaturized waveguides, respectively.

The results obtained may be extended to (i) different dielectric high-index materials, (ii) different meta-atoms geometries (e.g., rectangular pillars) and lattice types (e.g., triangular lattice) and, as a consequence, (iii) different frequency ranges including visible spectrum, which may be implemented, e.g., with silicon~\cite{schinke2015uncertainty} and titanium dioxide~\cite{sarkar2019hybridized} nanoparticles. The necessary material conditions to achieve a TE-TM polarization degeneracy of guided waves include the weak dispersion of permittivity and negligible absorption losses within the specified frequency range. Besides, the shape of nanoparticle in a meta-atom has to be adjusted in a way to achieve the close spectral position of electric and magnetic dipole-induced moments.


Based on the discovered principle, we use the platform of 1D meta-waveguides to create the planar polarizer of guided waves. The proposed concept allows to change the polarization of propagating guided waves at any frequency within the finite spectral range. We have numerically and experimentally investigated the quarter-wave-retardation near-field in-plane polarizer that acts as a linear-to-circular polarization transformer of guided waves.

It is worth noting that the discovered planar polarizer of guided waves is subwavelength in a sharp contrast to the typical waveguides and fibers. Generally, following the proposed principle one can create the polarizer with arbitrary phase delay by adjusting the 'anisotropic' part appropriately, namely the number and geometric parameters of cylinders. Even more, the proposed setup operates as a multifunctional frequency-dependent guided-wave polarizer with different phase delays depending on frequency, i.e. $\Delta k$ in 'anisotropic' part depends on frequency leading to the different total phase shifts within the same design of a polarizer. 

To conclude, we consider the results obtained as a significant milestone in planar polarization photonics. We believe the discovered platform of all-dielectric meta-waveguides with polarization-degenerate spectrum may lead to a plethora of applications in flat data transferring and processing devices, optical manipulation, antennas, polarization-sensitive pharmacology, and medicine.

\section{Materials and Methods}

\subsection{Numerical simulation and optimization \label{sec:num}}

The dispersion band diagram has been calculated using the frequency-domain Eigenmode Solver of CST Microwave Studio and MIT Photonics Band package~\cite{johnson2001block} for the 1D and 2D waveguides consisting of cylinders with $\varepsilon = 45$. In the first step, we analyzed the eigenmodes of the single cylinder to perform the initial selection of its diameter and height. In the second step, we aimed to minimize the $k$-DoPD in the vicinity of the first Brillouin zone ($k \approx k_B$). At the third step, we fixed the height as $h = 4.95$~mm and analyze the dependence of the $k$-DoPD ($\Delta k$) and DoL ($\Delta L$) on the period $a$ (Fig.~\ref{fig:degeneracy-vs-localization}). In the last step, we performed the final design optimization via the numerical simulations.

Then, using the Frequency Domain Solver in CST Microwave Studio, we perform the full-wave numerical calculation of guided waves in a finite metasurface and a chain consisting of 25$\times$25 and 51 ceramic cylinders, respectively, excited by the electric or magnetic point-like dipole source located above the central cylinder. The typical field distributions in $xy$-plane at $z=a/2$ are shown in Fig.~\ref{fig:2Dvs1D}c and \ref{fig:2Dvs1D}j. The dispersions of TM and TE guided waves have been recovered by applying the two-dimensional Fourier transform over the spatial distribution of normal components of the electric and magnetic fields, respectively~\cite{dockrey2016direct,yang2017hyperbolic,yermakov2018experimental,yermakov2021surface}, as it is shown in Figs.~\ref{fig:Figure3}a-\ref{fig:Figure3}c.

The planar near-field polarizer was numerically simulated in a similar way, while a dipole-like source was located 12~mm from the center of the edge cylinder along $x$-axis in order to excite the guided wave of a linear diagonal polarization ($S_2$ Stokes parameter). The distribution of the Stokes parameters (Figs.~\ref{fig:Stokes}c-\ref{fig:Stokes}d) and phase differences (Figs.~\ref{fig:polariz-phase}a-\ref{fig:polariz-phase}c) have been obtained in the $yz$-plane between the cylinders.

\subsection{Experimental sample and setup \label{sec:exp}}

The main element of our structure is a cylinder made of CaTiO$_3$-based ceramic composite with a diameter $d=5.20$~mm, height $h=4.95$~mm, and permittivity $\varepsilon=45$ (Figs.~\ref{fig:geom+setup}a-\ref{fig:geom+setup}b). The high-index cylindrical resonator exhibits both electric and magnetic Mie-like resonances~\cite{kruk2017functional}, which are studied in detail and engineered appropriately within this research. The periodic chain of cylinders (period $a=12$~mm) is located at the surface of a foam holder with near-unity permittivity (Fig.~\ref{fig:geom+setup}c).

We excite the TM (TE) guided waves via an electric dipole source realized as an open coaxial probe (a magnetic dipole source realized as a small loop antenna) located at the distance of 6~mm (half of the period) above the surface of the central cylinder of a chain (Fig.~\ref{fig:geom+setup}c). Then, we measure the spatial distribution of the normal components of the electric (magnetic) field using another electric-field probe (a magnetic field probe), which was realized in the same way as the source antenna, scanning along the structure in the $xy$-plane with a step of 3~mm using a precision 3-axis scanner. The applied procedure of the near-field measurements has been previously developed and verified~\cite{yermakov2018experimental,yermakov2021surface}. The signal obtained from the scanning electric/magnetic probes is recorded by the Rohde \& Schwarz ZVB20 vector network analyzer. The measurements were carried out in an anechoic chamber in the frequency range from 7.3 to 7.6~GHz with a spectral step of 50~MHz. The dispersions of TM and TE guided waves have been extracted using the two-dimensional Fourier transform~\cite{dockrey2016direct,yang2017hyperbolic,yermakov2018experimental,yermakov2021surface}, see Figs.~\ref{fig:Figure3}e-\ref{fig:Figure3}g.

The one-dimensional guided-wave polarizer represents a chain of different ceramic cylinders and consists of three parts. For the sake of simplicity, we label them as 'isotropic' and 'anisotropic'. The central 'anisotropic' part of the chain supporting the non-degenerate TE and TM eigenmodes with a non-zero wavevector difference ($\Delta k = |k_{\text{TE}} - k_{\text{TM}}|$) is sandwiched between the 'isotropic' chains of the above-mentioned cylinders supporting fully-degenerate eigenmodes spectrum with $\Delta k \approx 0$ (Fig.~\ref{fig:geom+setup}d). Here, the central part representing the ceramic cylinders with different sizes of height $h=5.4$~mm and $\Delta k \approx 5\cdot10^{-2}\pi/a$ plays a role of the polarization transformer (near-field analogue of a wave plate), while the side parts emulate the isotropic medium (near-field analogue of vacuum). The exciting magnetic loop is located in the vicinity of the left edge of the structure, and its plane is rotated by 45$^\circ$ with respect to $xy$-plane (Fig.~\ref{fig:geom+setup}d) in order to excite the guided wave with a linear under 45$^\circ$ polarization ($S_2 = 1$). The scanning loop was moving between the cylinders in the $yz$-plane before and after the in-plane polarizer. Moreover, at each measured point, the axis of the scanning magnetic probe (which is normal to the loop plane) was oriented along the $y$- and $z$-axis detecting the corresponding components of the magnetic field. We take 10 and 9 cylinders in 'isotropic' and each 'anisotropic' parts, respectively. The distribution of the Stokes parameters and phase differences are shown in Figs.~\ref{fig:Stokes}e-\ref{fig:Stokes}f and Figs.~\ref{fig:polariz-phase}e-\ref{fig:polariz-phase}g, respectively.





\bmhead{Acknowledgments}
The authors would like to thank A.~Sayanskiy, E.~Nenasheva and A.~Kalganov for their assistance in conducting the experiment.


\bibliography{sn-bibliography}

\end{document}